\documentclass[final,3p,times,twocolumn]{elsarticle}

\usepackage{lineno}
\modulolinenumbers[5]

\journal{Journal of .....}
\usepackage{graphicx}
\usepackage{dcolumn}
\usepackage{bm}
\usepackage{multirow} 
\usepackage{color}
\usepackage{graphicx} 
\usepackage{mathtools}
\usepackage{multirow}
\usepackage{caption}
\usepackage{amsmath} 
\usepackage{chemformula} 
\usepackage{csquotes} 
\biboptions{numbers,sort&compress}
\usepackage{subfig}
\usepackage{placeins}
\usepackage{cancel}
\usepackage[normalem]{ulem}
\usepackage{siunitx}
\usepackage{bm}
\usepackage[capitalise]{cleveref}
\usepackage{xr}
\usepackage{comment}
\usepackage{csvsimple}
\usepackage{breqn}
\usepackage{stfloats}

\newcommand{\eref}[1]{Eq.~(\ref{#1})}
\newcommand{\fref}[1]{Fig.~\ref{#1}}

\newcommand{\sref}[1]{Section~\ref{#1}}
\newcommand{\vm}[1]{\mathbf{#1}}

\newcommand{\bsym}[1]{\boldsymbol{#1}}

\makeatletter
\newcommand*{\addFileDependency}[1]{
  \typeout{(#1)}
  \@addtofilelist{#1}
  \IfFileExists{#1}{}{\typeout{No file #1.}}
}
\makeatother

\usepackage{amsmath, amssymb}
\usepackage{algorithm}
\usepackage{algpseudocode}
\usepackage{tcolorbox}
\usepackage{array}

\begin{document}

\begin{frontmatter}
\title{Graph Neural Network-Based Topology Optimization for Self-Supporting Structures in Additive Manufacturing}

\author[1,b]{Alireza Tabarraei\corref{mycorrespondingauthor}}
\ead{atabarra@charlotte.edu}
\author[1]{Saquib Ahmad Bhuiyan}
\cortext[mycorrespondingauthor]{Corresponding author}

\address[1]{Department of Mechanical Engineering and Engineering Science, The University of North Carolina at Charlotte, Charlotte, NC 28223, USA}
\address[b]{School of Data Science, The University of North Carolina at Charlotte, Charlotte, NC 28223, USA}

\begin{abstract}
This paper presents a machine learning-based framework for topology optimization of self-supporting structures, specifically tailored for additive manufacturing (AM). By employing a graph neural network (GNN) that acts as a neural field over the finite element mesh, the framework effectively learns and predicts continuous material distributions. An integrated AM filter ensures printability by eliminating unsupported overhangs, while the optimization process minimizes structural compliance under volume and stress constraints. The stress constraint is enforced using a differentiable $p$-norm aggregation of von Mises stress, promoting mechanical reliability in the optimized designs. A key advantage of the approach lies in its fully differentiable architecture, which leverages automatic differentiation throughout the optimization loop—eliminating the need for explicit sensitivity derivation for both the filter and the stress constraint. Numerical experiments demonstrate the ability of the framework to generate stress-constrained manufacturable topologies under various loading and boundary conditions, offering a practical pathway toward AM-ready high-performance designs with reduced post-processing requirements.
\end{abstract}

\begin{keyword}
\texttt Topology optimization \sep self Support structure \sep graph neural network \sep additive manufacturing \sep stress constraint.

\end{keyword}

\end{frontmatter}



\section{Introduction}\label{sec1}
Additive manufacturing, also known as 3D printing, has revolutionized the landscape of manufacturing, offering unprecedented design freedom and customization capabilities. Unlike traditional subtractive manufacturing methods, which involve cutting away material from a solid block, additive manufacturing builds objects layer by layer from digital designs. This layer-by-layer approach enables the fabrication of highly complex geometries with intricate internal structures, making it an ideal candidate for topology optimization.
Topology optimization aims to determine the optimal distribution of material within a given design domain to achieve desired performance objectives while satisfying various constraints. Traditionally, topology optimization has been utilized in engineering disciplines such as aerospace \cite{guanghui2020aerospace,zhu2016topology,hanush2022topology,leader2018high,rao2009topology}, automotive \cite{jankovics2019customization,cavazzuti2011high,sudin2014topology,yildiz2019topography}, and mechanical engineering \cite{tyflopoulos2022topology} to minimize weight, maximize stiffness, or enhance other performance metrics of structural components.
The synergy between topology optimization and additive manufacturing presents a compelling opportunity to push the boundaries of design and manufacturing. By leveraging the inherent capabilities of additive manufacturing processes, such as the ability to produce complex geometries without the need for traditional tooling, topology optimization can offer novel designs that were previously unattainable using conventional manufacturing techniques.

While AM offers numerous benefits, its layer-by-layer fabrication process presents distinct challenges, especially when producing geometries with overhanging features. 
The overhang angle, denoting the angle of inclination of a downward-facing surface with respect to the base plate, plays a crucial role in determining the efficacy of AM fabrication processes. If the overhang angle exceeds a critical angle, typically $45^\circ$, support structures will be required to prevent the collapse of newly printed layers. Though essential for achieving complex shapes and ensuring print quality, support structures have drawbacks. They increase material usage, elevating production costs and reducing efficiency. Additionally, removing support structures post-printing can be laborious and time-consuming, complicating the post-processing stage and extending lead times.  Furthermore, support structures can leave behind surface imperfections and blemishes on the printed part, diminishing the quality of the final product. These imperfections often necessitate meticulous post-processing techniques, further adding to the overall manufacturing time and cost.

Overcoming the challenges posed by overhangs has spurred the pursuit of topology optimization techniques aimed at eliminating these features from designs, thus obviating the necessity for support structures. Leary et al. \cite{leary2014optimal} addressed this issue by integrating a post-processing step into the density-based topology optimization method. This approach aimed to generate printable topological structures suitable for additive manufacturing. Essentially, this method involved incorporating features akin to support structures directly into the component design. However,  this method unavoidably alters the target weight and compromises the optimality of the original optimized designs. 
Gaynor and Guest \cite{gaynor2016topology} introduced a wedge-shaped filter technique, wherein the density value at a particular point is linked with the filter region below to depict the support condition. 
This technique was later extended into three dimensions by Johnson and Gaynor \cite{johnson2018three}, ensuring that a structural point cannot exist unless neighboring points adequately support it.
Qian \cite{qian2017undercut} investigated a projection-based method, utilizing the projected overhang length to mitigate overhang features. Building upon this, Wang and Qian \cite{wang2020simultaneous} extended the approach to optimize both the build orientation and overhang angle. 
Kuo et al. \cite{kuo2018support} introduced a framework that utilizes a cost-based formulation to generate designs characterized by easily removable supports and improved surface profiles. To foster the creation of additive manufacturing  friendly designs, Mirzendehdel and Suresh \cite{mirzendehdel2016support} regulated the volume of support structures based on topological sensitivity.
Langelaar \cite{langelaar2016topology, langelaar2017additive} developed a layer--by--layer overhang filter to remove the issue of overhangs. He used this filter to simultaneously optimize part topology, support structure layout, and build orientation \cite{langelaar2018combined}. 
van de Ven et al. \cite{van2021overhang} devised an overhang filter utilizing front propagation, enabling the detection of overhang regions using an anisotropic speed function specifically tailored for density-based topology optimization. This approach offers the advantage of accommodating an unstructured mesh, which is well-suited for initial design domains featuring irregularities such as holes and curved surfaces. However, it comes with a higher computational cost compared to the layer-by-layer filtering method devised by Langelaar \cite{langelaar2016topology, langelaar2017additive}.

Equally important to manufacturability is the mechanical performance of AM-produced structures under applied loading. To ensure functional reliability, AM components must be designed to resist stress concentrations and avoid excessive deformation or failure during service. To address this, recent topology optimization frameworks have begun incorporating stress constraints in addition to geometric printability criteria. This dual consideration enables the generation of designs that are not only printable but also mechanically robust. Notably, the use of differentiable global stress formulations—such as $p$-norm  aggregations of element-wise von Mises stress—has made it possible to enforce strength constraints efficiently within gradient-based optimization pipelines, allowing for smooth and scalable integration of structural performance requirements during design.

Despite these advances, topology optimization remains computationally expensive due to the large number of design variables and the iterative nature of finite element analysis (FEA). Various strategies have been proposed to accelerate the process, including GPU-accelerated solvers~\cite{martinez2017efficient} and parallel computing frameworks~\cite{aage2015topology}, yet the high computational cost remains a central challenge.
To address this, machine learning (ML) has emerged as a powerful tool to accelerate and automate topology optimization. Generative approaches such as GANs~\cite{parrott2023multidisciplinary} and diffusion models~\cite{maze2023diffusion,zhang2025research, lutheran2025latent} have been used to synthesize optimal topologies directly from boundary condition images. Although these methods provide rapid predictions once trained, they often require large datasets and struggle to integrate complex manufacturing constraints such as overhang control. To reduce the reliance on data, Chi et al.~\cite{chi2021universal} proposed a framework that leverages historical optimization data to learn the relationship between design variables and sensitivities, offering a more interpretable and data-efficient alternative.

Another promising direction involves the use of neural fields to represent material distributions within the design domain \cite{chandrasekhar2021tounn,  shishir2023topology, tabarraei2025variational}. For example, neural networks have been integrated with FEA to predict element-wise material layouts while relying on conventional solvers to compute structural responses and guide optimization. This hybrid approach has been successfully extended to more advanced cases, including multi-material systems and composite structures, demonstrating the flexibility of neural models across a broad spectrum of topology optimization problems~\cite{chandrasekhar2021multi,shishir2024multi}.

In this work, we propose a differentiable, simulation-integrated framework for topology optimization using neural fields built upon graph neural networks. The finite element mesh is represented as a graph, where each node corresponds to an element and edges encode neighborhood relationships. Each node is enriched with spatial information using Fourier feature mappings, which embed positional coordinates into a higher-dimensional space to capture both local and global spatial dependencies. A Chebyshev-based GNN then predicts the material distribution as a continuous density field over the design domain.
These predicted densities are passed to a finite element solver, which evaluates structural performance through compliance and von Mises stress computations. To ensure printability, we incorporate a differentiable AM filter that suppresses unsupported overhangs during the forward pass. The loss function combines compliance, material volume, and a smooth global p-norm stress constraint, guiding the network toward mechanically robust, AM-compatible designs.

Leveraging automatic differentiation, this framework enables end-to-end optimization without requiring manual gradient derivations, even when incorporating complex manufacturing and mechanical constraints. The integration of neural fields, Fourier encoding, graph-based learning, and simulation-driven evaluation presents a scalable and flexible approach for structural design under additive manufacturing constraints.

\section{Problem statement }
The solid isotropic material with penalization (SIMP) method  \cite{bendsoe1989optimal,sigmund2007morphology,andreassen2011efficient,zhou1991coc} is a widely adopted approach in the field of topology optimization, which is a mathematical technique used to design optimal structures and materials. The SIMP method utilizes a material interpolation scheme that allows for a smooth transition between material and void (empty space) within the design domain. This is achieved by assigning each element in the design domain a density variable that can vary between 0 (void) and 1 (solid material). The mechanical properties of each element, such as Young's modulus, are then interpolated between those of the solid material and an almost negligible value, corresponding to the void, based on the element’s density
\begin{equation}
E_e(\rho_e) = E_{min} + \rho_e^p (E_0 - E_{min})
\end{equation}
where $p$ is the penalization parameter, $\rho_e$ is the printing density of element $e$, $E_{min}$ is a negligible value representing the Young's modulus of the void elements and 
$E_0$ is the Young's modulus of the solid materials. 

Various objective functions and constraints have been used for topology optimization purposes.  One of the most common objective functions used in this context is the minimization of the compliance function, which is closely related to maximizing the stiffness of the structure.
A volume constraint is often imposed to limit the amount of material used, corresponding to controlling the structure's weight.
The standard form of the topology optimization problem with these objectives and constraints can be mathematically stated as
\begin{equation}
\begin{aligned}
& \underset{\vm{b}}{\text{minimize}} \quad C =  \mathbf{f}^T \mathbf{u(\bsym{\rho})}  \\
&\begin{aligned}
&\textrm{subject to} 
&\left\{
\begin{aligned}
&V(\bsym{\rho}) \le V_{\mathrm{max}}  \\
&\mathbf{K}(\bsym{\rho})\mathbf{u}(\bsym{\rho}) = \vm{f} \\
& 0 \le\vm{b} \le 1
\end{aligned}
\right.
\end{aligned}
\end{aligned}
\end{equation}
In this context, $\vm{K}$, $\vm{u}$, and $\vm{f}$ denote the stiffness matrix, displacement vector, and load vector of the finite element system, respectively. $V$ and $V_{\mathrm{total}}$ refer to the actual and permissible volumes of the printed part. It is important to note that both the compliance $C$ and the volume $V$ are assessed based on the printed density field $\bsym{\rho}$, whereas the optimization process adjusts the blueprint field $\vm{b}$. 

\section{The overhang filter}\label{amFilter}
\begin{figure}[tb]
  \centering
\includegraphics[width = 0.95\linewidth, trim=0 5pt 0 0,clip]{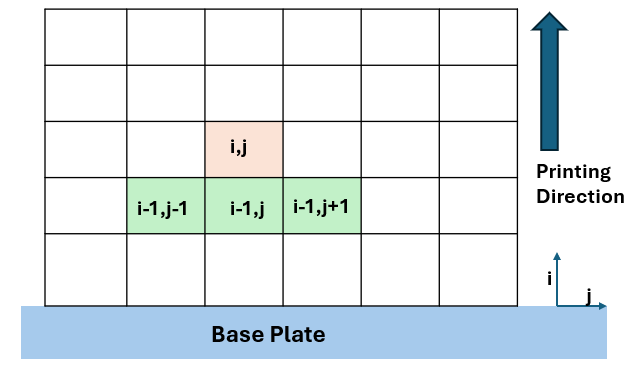}
  \caption{Schematic diagram of AM filter}
    \label{amfilter}
\end{figure}
In this study, we utilize the layerwise filter operation developed in Reference \cite{langelaar2016topology}  to address the challenges of overhang constraints in AM processes during topology optimization. This filter transforms an initial blueprint design into a geometry that is printable without the need for additional support. It is applied to a uniform regular mesh, which is common in early-stage topology optimization. Consider a rectangular domain discretized by $n_i\times n_j$ elements, where the printing direction is in the vertical direction. 
For each element at location $(i,j)$, the supporting region $S_{i,j}$, shown in \fref{amfilter}, comprises the elements at positions  
$(j-1)$ to $(j+1)$ of the below layer $(i-1)$.
An element at position $(i,j)$
must receive sufficient support from these elements in layer $i-1$ to be considered printable.
All elements in the first layer are automatically deemed printable. However, for the subsequent layers, the printed 
density $\rho_{i,j}$ of an element cannot exceed
the maximum printed density $E_{i,j}$ observed within its supporting region $S_{i,j}$. This can be written mathematically as
\begin{subequations}
    \begin{align}
        \rho_{i,j} &= \min(b_{i,j}, E_{i,j}) \\
        E_{i,j} &= \max(\rho_{i-1, j-1}, \rho_{i-1, j}, \rho_{i-1, j+1})
    \end{align}
\end{subequations}
where $b_{i,j}$ is the blueprint density of element $(i,j)$.

Since the minimum and maximum functions are not differentiable, they are replaced with the following functions

\begin{subequations} \label{approx}
\begin{align}
\rho_{i,j} &\approx S(b_{i,j}, E_{i,j}) \nonumber \\
&= \frac{1}{2} \left( 
b_{i,j} + E_{i,j} 
- \left( (b_{i,j} - E_{i,j})^2 + \epsilon \right)^{0.5} 
+ \sqrt{\epsilon} \right) \label{approx:a} \\[6pt]
E_{i,j} &\approx \left( 
\rho_{i-1,j-1}^P + \rho_{i-1,j}^P + \rho_{i-1,j+1}^P 
\right)^{1/Q} \label{approx:b} \\[6pt]
Q &= P + \frac{\log n_s}{\log \epsilon_0} \label{approx:c}
\end{align}
\end{subequations}

where $\epsilon$ and $P$ 
dictate the precision and smoothness of the approximation. As 
$\epsilon$ approaches 0 and $P$ approaches infinity, the approximations converge to the exact minimum and maximum functions. In \eref{approx}, $n_s$ represents the number of elements within the support domain ($n_s 
 = 3$ in this paper), and we set $\epsilon_0 = 0.5$ as the default value. 

 \subsection{Sensitivity analysis}\label{amSensitivity}
To address the optimization problem using gradient-based methods, it is crucial to determine the sensitivity of each response function $g$ concerning the blueprint density $\mathbf{b}$. The response function $g$ is dependent on the printed density $\boldsymbol{\rho}$, which itself is derived from the blueprint density $\mathbf{b}$ as described in Equation \ref{approx}. Therefore, the relationship can be expressed as $g(\boldsymbol{\rho}(\mathbf{b}))$. The response functions are 
calculated using the finite element method. 
Using the chain's rule, the derivative of the response function with respect to the blueprint 
density $\vm{b}$ can be written as
\begin{equation}
    \frac{\partial g}{\partial \vm{b}} = \frac{\partial g}{\partial \bsym{\rho}}\frac{\partial \bsym{\rho}}{\partial \vm{b}}
\end{equation}
Based on \eref{approx}, the printed density field $\boldsymbol{\rho}_k$ of layer $k$,  evolves layer by layer, depending on both the blueprint density $b_k$ and the previously printed layer $\boldsymbol{\rho}_{k-1}$. This relationship can be expressed as
\begin{equation}
\boldsymbol{\rho}_k = S(\vm{b}_k, \bsym{\rho}_{k-1}), \quad k = 2, \dots, n.
\end{equation}
where $n$ is the number of layers. For the first layer the printed density is the same as the blueprint density, i.e. $\bsym{\rho}_1 = \bsym{b}_1$. 

The adjoint sensitivity analysis can be used to compute the sensitivities. For this 
purpose, the augmented response function incorporating Lagrange multipliers $\bsym{\lambda}_k$
can be defined as
\begin{equation}
\tilde{g} = g (\bsym{\rho}(\vm{b})) + \sum_{k=2}^{n} \bsym{\lambda}_k^T (S(\vm{b}_k, \bsym{\rho}_{k-1}) - \vm{p}_k),
\end{equation}
For the base layer ($k = 1$), $\bsym{\rho}_1 = \vm{b}_1$ hence $\frac{\partial g}{\partial \vm{b}_1} = \frac{\partial g}{\partial \bsym{\rho}_1}$. For subsequent layers, differentiating $\tilde{g}$ with respect to $\vm{b}_m$ results in
\begin{align}
\frac{\partial \tilde{g}}{\partial \vm{b}_m} 
&= \sum_{k=2}^{n} \left\{ \frac{\partial g}{\partial \bsym{\rho}_k} \frac{\partial \bsym{\rho}_k}{\partial \vm{b}_m} 
+ \lambda_k^T \left( \frac{\partial S}{\partial \vm{b}_m} \delta_{km} \right. \right. \notag \\
&\quad \left. \left. + \frac{\partial S}{\partial \bsym{\rho}_{k-1}} \frac{\partial \bsym{\rho}_{k-1}}{\partial \vm{b}_m} 
- \frac{\partial \bsym{\rho}_k}{\partial \vm{b}_m} \right) \right\}
\end{align}

where \( \delta_{km} \) denotes the Kronecker delta, and \( 1<m\leq n \). Since printed densities only depend on blueprint densities in underlying layers, terms in the summations with \( k<m \) vanish. This allows the previous equation  to be rewritten as:
\begin{align}
\frac{\partial \tilde{g}}{\partial \mathbf{b}_m} 
&= \frac{\partial g}{\partial \boldsymbol{\rho}_m} \frac{\partial \boldsymbol{\rho}_m}{\partial \mathbf{b}_m} 
+ \sum_{k=m+1}^{n} \bigg\{ \left( \frac{\partial g}{\partial \boldsymbol{\rho}_k} - \boldsymbol{\lambda}_k^T \right) \frac{\partial \boldsymbol{\rho}_k}{\partial \mathbf{b}_m} \notag \\
&\quad + \boldsymbol{\lambda}_k^T \frac{\partial S}{\partial \boldsymbol{\rho}_{k-1}} \frac{\partial \boldsymbol{\rho}_{k-1}}{\partial \mathbf{b}_m} \bigg\}
\end{align}

By expanding the summation and noting that 
 $\frac{\partial \boldsymbol{\rho}_m}{\partial \mathbf{b}_m} = \frac{\partial S}{\partial \mathbf{b}_m}$, the previous equation can be written as 
\begin{align}
    \frac{\partial \tilde{g}}{\partial \mathbf{b}_m} &= 
    \left(\frac{\partial g}{\partial \boldsymbol{\rho}_m} + \boldsymbol{\lambda}_{m+1}^T \frac{\partial S}{\partial \boldsymbol{\rho}_m} \right) 
    \frac{\partial S}{\partial \mathbf{b}_m} \nonumber \\
    &\quad + \left(\frac{\partial g}{\partial \boldsymbol{\rho}_{m+1}} - \boldsymbol{\lambda}_{m+1}^T 
    + \boldsymbol{\lambda}_{m+2}^T \frac{\partial S}{\partial \boldsymbol{\rho}_{m+1}} \right) 
    \frac{\partial \boldsymbol{\rho}_{m+1}}{\partial \mathbf{b}_m} \nonumber \\
    &\quad + \left(\frac{\partial g}{\partial \boldsymbol{\rho}_{m+2}} - \boldsymbol{\lambda}_{m+2}^T 
    + \boldsymbol{\lambda}_{m+3}^T \frac{\partial S}{\partial \boldsymbol{\rho}_{m+2}} \right) 
    \frac{\partial \boldsymbol{\rho}_{m+2}}{\partial \mathbf{b}_m} \nonumber \\
    &\quad + \dots \nonumber \\
    &\quad + \left(\frac{\partial g}{\partial \boldsymbol{\rho}_{n-1}} - \boldsymbol{\lambda}_{n-1}^T 
    + \boldsymbol{\lambda}_{n}^T \frac{\partial S}{\partial \boldsymbol{\rho}_{n-1}} \right) 
    \frac{\partial \boldsymbol{\rho}_{n-1}}{\partial \mathbf{b}_m} \nonumber \\
    &\quad + \left(\frac{\partial g}{\partial \boldsymbol{\rho}_{n}} - \boldsymbol{\lambda}_{n}^T \right) 
    \frac{\partial \boldsymbol{\rho}_{n}}{\partial \mathbf{b}_m}
\end{align}
which in compact form can be written as 
\begin{align}\label{sens1}
    \frac{\partial \tilde{g}}{\partial \mathbf{b}_m} &= 
    \left( \frac{\partial g}{\partial \boldsymbol{\rho}_m} 
    + \boldsymbol{\lambda}_{m+1}^T \frac{\partial S}{\partial \boldsymbol{\rho}_m} \right) 
    \frac{\partial S}{\partial \mathbf{b}_m} \nonumber \\
    &\quad + \sum_{k=m+1}^{n_z-1} \left( \frac{\partial g}{\partial \boldsymbol{\rho}_k} 
    - \boldsymbol{\lambda}_k^T + \boldsymbol{\lambda}_{k+1}^T \frac{\partial S}{\partial \boldsymbol{\rho}_k} \right) 
    \frac{\partial \boldsymbol{\rho}_k}{\partial \mathbf{b}_m} \nonumber \\
    &\quad + \left( \frac{\partial g}{\partial \boldsymbol{\rho}_{n}} 
    - \boldsymbol{\lambda}_{n}^T \right) \frac{\partial \boldsymbol{\rho}_{n}}{\partial \mathbf{b}_m}
\end{align}
Based on \eref{sens1}, computing $\frac{\partial \boldsymbol{\rho}_k}{\partial \mathbf{b}_m}$ can be avoided if the Lagrange multipliers are chosen as
\begin{align}\label{lambda}
    \bsym{\lambda}_k^T &= 
    \begin{cases} 
        \frac{\partial g}{\partial \bsym{\rho}_k} + \bsym{\lambda}_{k+1}^T \frac{\partial S}{\partial \bsym{\rho}_k}, & \quad \text{for} \ \ 1<k<n \\[8pt]
        \frac{\partial g}{\partial \bsym{\rho}_k}, & \quad \text{for} \ \ k= n
    \end{cases}
\end{align}
This recursive relationship enables the computation of Lagrange multipliers by iterating through the domain from the top layer to the bottom layer.

\section{Topology Optimization with Stress and Additive Manufacturing Constraints}

In structural applications of additive manufacturing, ensuring that optimized topologies are both printable and mechanically reliable is crucial. While printability is addressed through the AM filter introduced earlier, this section focuses on enforcing structural integrity under applied loads by incorporating stress control into the optimization process.

Stress constraints are enforced using a global $p$-norm aggregation of von Mises stress across the domain \cite{shishir2023topology,kiyono2016new,holmberg2013stress,lee2016novel,fan2019evolutionary,zheng2023stress,yaghoobi2025introducing}. This approach provides a smooth and differentiable approximation of the maximum stress, enabling efficient integration into gradient-based optimization routines. The stress field is computed via linear elastic analysis under plane stress conditions, using the filtered material density to assemble the system stiffness.

\subsection{Stress Computation and Aggregation}

The displacement field $\vm{u}$ is obtained by solving the finite element equilibrium equation
\begin{equation}
\vm{K}(\bsym{\rho}) \, \vm{u} = \vm{f}
\end{equation}
where $\vm{K}$ is the global stiffness matrix assembled using the printed density $\bsym{\rho}$. For each element $e$, the stress tensor $\bsym{\sigma}_e$ is evaluated at its centroid using
\begin{equation}
\bsym{\sigma}_e = \vm{D}_e \, \vm{B} \, \vm{u}_e
\end{equation}
The von Mises stress is then computed as
\begin{equation}
\sigma^{e}_{\text{vm}} = \sqrt{\sigma_{xx}^2 + \sigma_{yy}^2 - \sigma_{xx} \sigma_{yy} + 3\sigma_{xy}^2}
\end{equation}
To maintain consistency with the material interpolation, we scale the stress by $\sqrt{E_e}$ to reflect the stiffness-weighted contribution of each element
\begin{equation}
\sigma^{e}_{\text{vm}} \leftarrow \sigma^{e}_{\text{vm}} \cdot \sqrt{E_e}
\end{equation}

A smooth approximation of the maximum stress is formulated using a global $p$-norm:
\begin{equation}
\sigma_{\text{PN}} = \left( \frac{1}{N} \sum_{e=1}^{N} \left( \frac{\sigma^{e}_{\text{vm}}}{\bar{\sigma}} \right)^p \right)^{1/p} - 1
\end{equation}
where $\bar{\sigma}$ is the allowable stress and $N$ is the number of finite elements. The global constraint $\sigma_{\text{PN}} \leq 0$ ensures that the structure remains within the permissible stress range without requiring element-wise constraints.

\subsection{Gradient Computation via Adjoint Method}\label{pnormSens}

The derivative of the global stress constraint with respect to the design variable $b_j$ is evaluated using the chain rule
\begin{equation}
\frac{\partial \sigma_{\text{PN}}}{\partial b_j} =
\left( \sum_{e=1}^{N} \left( \frac{\sigma^{e}_{\text{vm}}}{\bar{\sigma}} \right)^p \right)^{\frac{1}{p} - 1}
\cdot \sum_{e=1}^{N} \frac{p}{\bar{\sigma}^p} \left( \sigma^{e}_{\text{vm}} \right)^{p-1} \cdot \frac{\partial \sigma^{e}_{\text{vm}}}{\partial b_j}
\end{equation}
The element-wise stress sensitivity is given by
\begin{equation}
\frac{\partial \sigma^{e}_{\text{vm}}}{\partial b_j} =
\frac{1}{\sigma^{e}_{\text{vm}}} \, \bsym{\sigma}_e^\top \, \frac{\partial \bsym{\sigma}_e}{\partial b_j}
\end{equation}
where $\partial \bsym{\sigma}_e/\partial b_j$ involves both the stiffness interpolation and the displacement sensitivity.

The displacement sensitivity is computed using the adjoint method. Defining the adjoint vector $\bsym{\lambda}$ as the solution of
\begin{equation}
\vm{K} \bsym{\lambda} = - \sum_{e} \left( \frac{\partial \sigma^{e}_{\text{vm}}}{\partial \vm{u}_e} \right)^\top \cdot \frac{\partial \sigma_{\text{PN}}}{\partial \sigma^{e}_{\text{vm}}}
\end{equation}
the total derivative is then assembled as
\begin{equation}
\frac{\partial \sigma_{\text{PN}}}{\partial b_j} =
\bsym{\lambda}^\top \, \frac{\partial \vm{K}}{\partial b_j} \, \vm{u}
+ \sum_{e} \frac{\partial \sigma_{\text{PN}}}{\partial \sigma^{e}_{\text{vm}}} \cdot \left(
\frac{\partial \bsym{\sigma}_e}{\partial E_e} \cdot \frac{\partial E_e}{\partial b_j}
\right)
\end{equation}
The first term reflects the influence of density on the structural response, while the second captures the direct effect of stiffness variation on local stress magnitudes.

\section{Remark on Sensitivities and Computational Burden}
While the sensitivity analyses for the AM filtering operation and the $p$-norm stress constraint have been derived and discussed for completeness in \sref{amSensitivity} and \sref{pnormSens}, it is important to emphasize that these gradients are not required in our proposed framework. In conventional topology optimization pipelines, these sensitivity terms are essential for enabling gradient-based updates through hand-crafted adjoint formulations. However, the presence of nested max/min and norm operations—particularly in layerwise filtering and stress aggregation—makes the exact gradient computation computationally intensive and algorithmically intricate, especially when scaling to high-resolution 3D problems.

Our approach circumvents this challenge by leveraging a fully differentiable, end-to-end neural field architecture. Since the material distribution is predicted via a graph neural network conditioned on spatially encoded input, and the entire pipeline (including the finite element solver and AM filter) is implemented in a differentiable programming environment, gradients are computed automatically via backpropagation. Thus, the explicit calculation of complex analytical sensitivities becomes unnecessary. The inclusion of these derivations in this paper is intended to highlight the underlying challenges of conventional formulations and to underscore the advantages offered by a neural field-based topology optimization framework.


\section{Graph convolution neural network for topology optimization}
\begin{figure*}[t!]
	\centering
	\includegraphics[width = 0.9578\linewidth, height=0.43\linewidth]{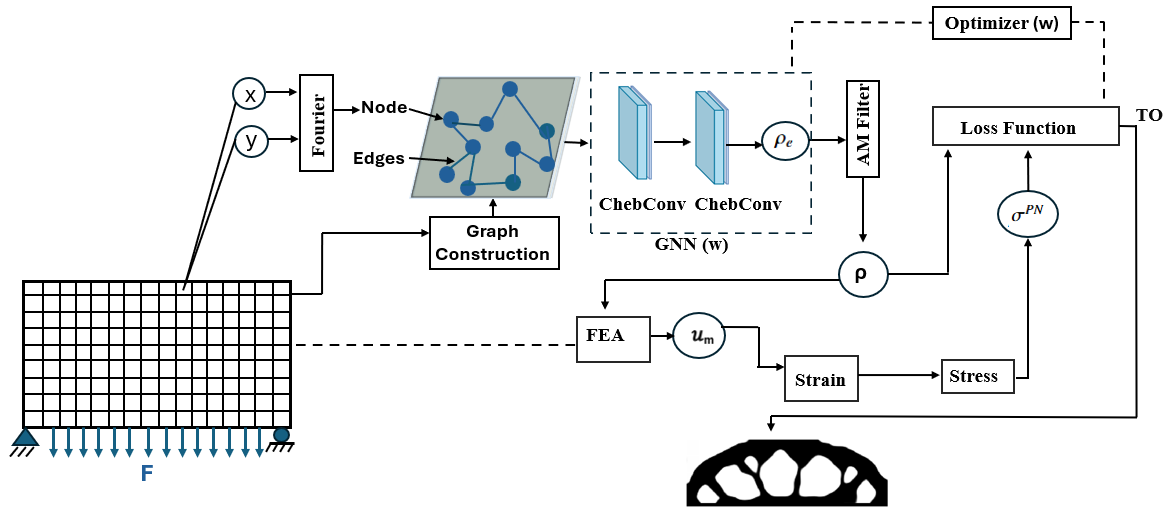} 
	\caption{Illustration of Proposed Framework.}
    \label{framework}
\end{figure*} 
\begin{figure*}[t!]
	\centering
	\includegraphics[width = 0.8278\linewidth, height=0.33\linewidth]{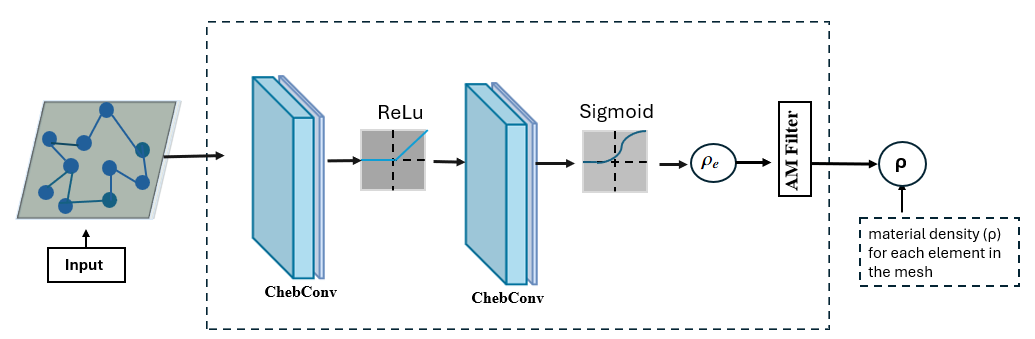} 
	\caption{Overview of the graph neural network architecture used for topology optimization.}
    \label{framework_GCN}
\end{figure*} 

A graph is composed of nodes, each characterized by a feature vector $\mathbf{h}$, and edges that define the connections among them. The neighborhood $\mathcal{N}(v)$ of a node $v$ consists of all nodes directly connected to it. GNNs operate through two core operations: message passing and aggregation. During message passing, each node transmits information to its neighbors, which is then aggregated to update the node's features. This mechanism enables GNNs to encode localized geometric and topological relationships into node embeddings.

In our work, we employ a graph convolutional network (GCN) to predict the blueprint density \( b_e \) for each finite element. GCNs offer a natural fit for unstructured domains and non-Euclidean geometries due to their inherent flexibility in processing arbitrary graph topologies \cite{bhuiyan2025graph}. Unlike conventional CNNs restricted to regular grids, GCNs generalize convolution operations to graph-structured data, making them ideal for encoding spatially distributed fields such as material densities over FEM meshes.

The GCN architecture consists of multiple graph convolution layers that iteratively update node features. A general graph convolution operation can be expressed as:
\begin{equation}
\mathbf{h}_v^l = \gamma^l \left( \mathbf{h}_v^{l-1}, \sigma^l \left( \left\{ \phi^l(\mathbf{h}_u^{l-1}) : u \in \mathcal{N}(v) \right\} \right) \right)
\end{equation}
where $\mathbf{h}_v^{l-1}$ and $\mathbf{h}_v^{l}$ are the feature vectors of node $v$ at layers $l{-}1$ and $l$, respectively, $\phi$ is a message function applied to neighboring features, $\sigma$ is an aggregation operator (e.g., mean or sum), and $\gamma$ is the update function combining local and neighbor information.

We adopt the Chebyshev spectral graph convolution operator~\cite{he2022convolutional}, which approximates spectral filters using Chebyshev polynomials to avoid costly eigendecompositions. The node update rule is defined as:
\begin{equation}
\mathbf{h}_v^l = \text{ReLU} \left( \sum_{k=0}^K \boldsymbol{\theta}_k \, \mathbf{T}_k(\tilde{\mathbf{L}}) \, \mathbf{h}^{l-1} + \mathbf{b} \right)
\end{equation}
where \( \mathbf{T}_k(\tilde{\mathbf{L}}) \) is the \( k \)-th Chebyshev polynomial of the scaled graph Laplacian \( \tilde{\mathbf{L}} \), \( \boldsymbol{\theta}_k \) are trainable weights, and \( \mathbf{b} \) is the bias vector. The ReLU activation introduces nonlinearity.

The Chebyshev polynomials of the first kind are recursively defined as:
\begin{align*}
\mathbf{T}_0(x) &= 1, \\
\mathbf{T}_1(x) &= x, \\
\mathbf{T}_k(x) &= 2x\mathbf{T}_{k-1}(x) - \mathbf{T}_{k-2}(x)
\end{align*}

The scaled Laplacian \( \tilde{\mathbf{L}} \) is computed as:
\[
\tilde{\mathbf{L}} = \frac{2\mathbf{L}}{\lambda_{\text{max}}} - \mathbf{I}
\]
where \( \mathbf{L} = \mathbf{D}^{-1/2}\mathbf{A}\mathbf{D}^{-1/2} - \mathbf{I} \) is the normalized Laplacian, \( \mathbf{A} \) is the adjacency matrix, \( \mathbf{D} \) the degree matrix, and \( \lambda_{\text{max}} \) the largest eigenvalue of \( \mathbf{L} \). This scaling ensures numerical stability and supports effective learning with high-order polynomial filters.

\section{Proposed framework}

The proposed framework integrates a graph-based material prediction model with a physics-informed topology optimization pipeline that enforces structural and AM constraints through automatic differentiation. The framework, as illustrated in \fref{framework}, integrates three primary modules in a fully differentiable loop; graph-based material density prediction, finite element simulation, and physics-informed loss computation. This end-to-end approach eliminates the need for manual sensitivity derivation and allows seamless incorporation of manufacturability constraints via automatic differentiation.

The design domain is discretized using a regular finite element mesh, where each element is treated as a node in a graph. Spatial relationships between elements are represented as edges connecting neighboring nodes. For each node $v$, the input feature consists of its centroid coordinates, which are encoded using a Fourier mapping $\gamma(\mathbf{x}_v) \in \mathbb{R}^{2m}$ to enrich the spatial representation:

\begin{equation}
\gamma(\mathbf{x}_v) = \left[ \sin(2\pi \mathbf{B} \mathbf{x}_v), \; \cos(2\pi \mathbf{B} \mathbf{x}_v) \right]
\tag{33}
\end{equation}

Here, $\mathbf{x}_v \in \mathbb{R}^2$ is the spatial coordinate of node $v$, and $\mathbf{B} \in \mathbb{R}^{m \times 2}$ is a matrix of random frequencies. This graph is passed to a Chebyshev-based Graph Convolutional Network (ChebNet) that propagates feature information across the mesh. The GNN comprises multiple graph convolution layers that transform and aggregate node features, ultimately predicting a pseudo-density $\rho_e \in [0, 1]$ for each element $e$. These predicted densities represent the material layout within the design domain.

To ensure manufacturability, the pseudo-densities are passed through a layer-by-layer AM filter described in \sref{amFilter} that enforces overhang constraints by restricting upward growth of unsupported elements. This filter operates iteratively along the print direction and outputs the printed density $\rho_e$, which is then used in the physics computations.
The filtered density $\rho_e$ is used to assemble the global stiffness matrix $\mathbf{K}(\rho)$ through SIMP interpolation of material properties. The mechanical equilibrium equation
is solved to compute the global displacement field $\mathbf{u}$. Element-wise displacements $\mathbf{u}_e$ are extracted to evaluate strains and stresses. Von Mises stress is calculated per element and then aggregated via a $p$-norm  formulation to obtain the global stress measure $\sigma_{\text{PN}}$. 

The predicted material distribution is optimized using a composite loss function that includes terms for structural performance, material efficiency, and stress control. The loss is defined as
\begin{equation}
\begin{aligned}
L(\mathbf{w}) = & \, \frac{J(\rho_e)}{J_0} 
+ \alpha \left( \frac{\sum_e \rho_e v_e}{V^\star} - 1 \right)^2 \\
& + \gamma \sum_{i=1}^{n_c} \left( 
\left( \frac{1}{N_i} \sum_{a \in \Omega_i} \left( \frac{\sigma_a^{vM}(x)}{\bar{\sigma}} \right)^p \right)^{\frac{1}{p}} - 1 
\right)^2
\end{aligned}
\end{equation}
where $\alpha$ and $\gamma$ are constraint penalty parameters, $\bar{\sigma}$ is the allowable stress, and $J_0$ represents the initial compliance of the system, used here as a scaling factor for the compliance.

The entire pipeline is implemented in PyTorch, and all operations are differentiable. During each iteration, the GNN predicts pseudo-densities, the filtered densities are passed to the FEA module, and the resulting compliance and stress are used to compute the loss. The gradient of this loss with respect to the GNN weights 
$w$ is computed via automatic differentiation and used to update the network using the Adam optimizer \cite{kingma2014adam,zhang2018improved}. This framework enables fully automated topology optimization that enforces complex stress and manufacturing constraints, producing manufacturable structures without the need for explicitly coded sensitivity analysis.
\section{Results and Discussion}
\begin{figure}[tb]
  \centering
\includegraphics[width = 0.96\linewidth, trim=0 5pt 0 0,clip]{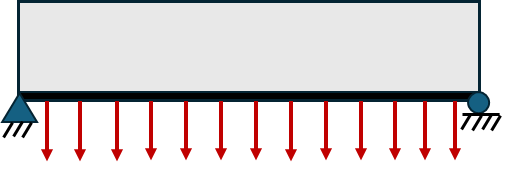}
  \caption{A simply supported beam under distributed loading.}
    \label{simpBeam}
\end{figure}
\begin{figure*}[tb]
  \centering
  \subfloat[Compliance = 30.00e3]{\includegraphics[width = 0.3\linewidth, trim=0 5pt 0 0,clip]{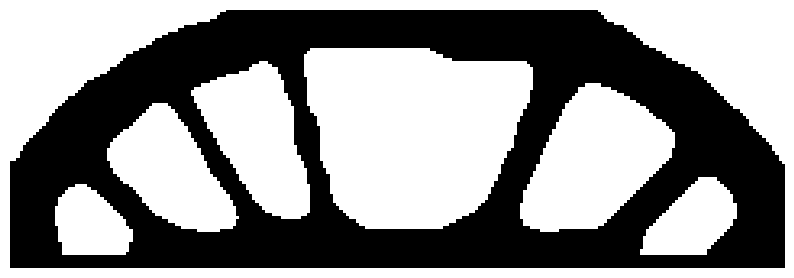}} 
  \hspace{0.4cm} 
  \subfloat[Compliance = 33.31e3]{\includegraphics[width = 0.3\linewidth, trim=0 5pt 0 0,clip]{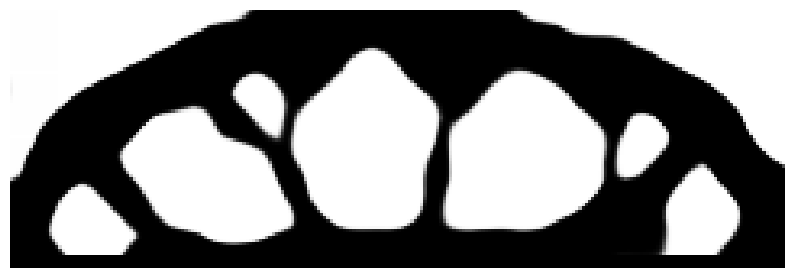}}
  \hspace{.4cm} 
  \subfloat[Compliance = 35.65e3]
  {\includegraphics[width = 0.3\linewidth, trim=0 5pt 0 0,clip]{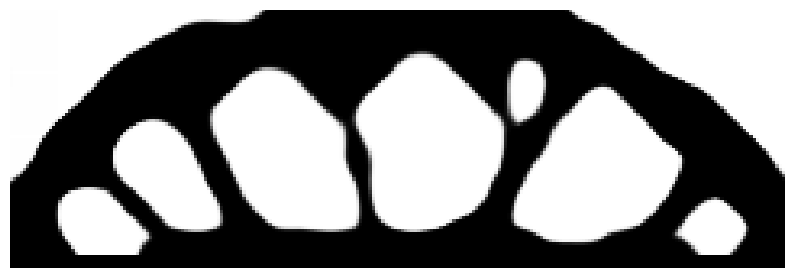}}
  \hspace{0.4cm}
  \caption{Designed topology for a simply supported beam under
a distributed load  a) without AM filter, b) with AM filter c) with AM filter + stress constraint. }
    \label{with_stress_constraints}
\end{figure*}

\begin{figure*}[tb!]
	\centering
	\includegraphics[width = 0.9078\linewidth, trim=0 0 0 0,clip]{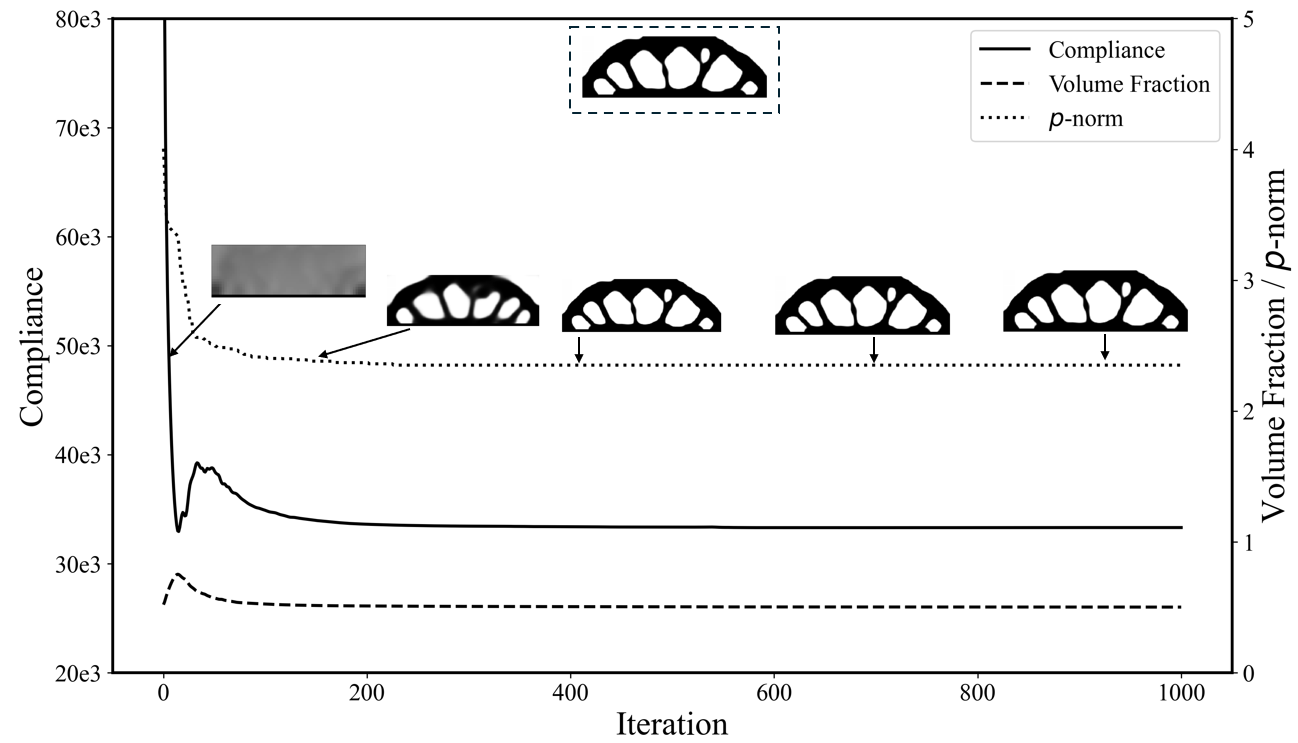} 
	\qquad	
\caption{Evolution of compliance, P-Norm volume fraction over iterations for simply supported under a distributed load .}
    \label{eval_iter_Pnorm}
\end{figure*}

  
\begin{figure}[tb!]
	\centering
	\includegraphics[width = 0.9978\linewidth, trim=0 0 0 0,clip]{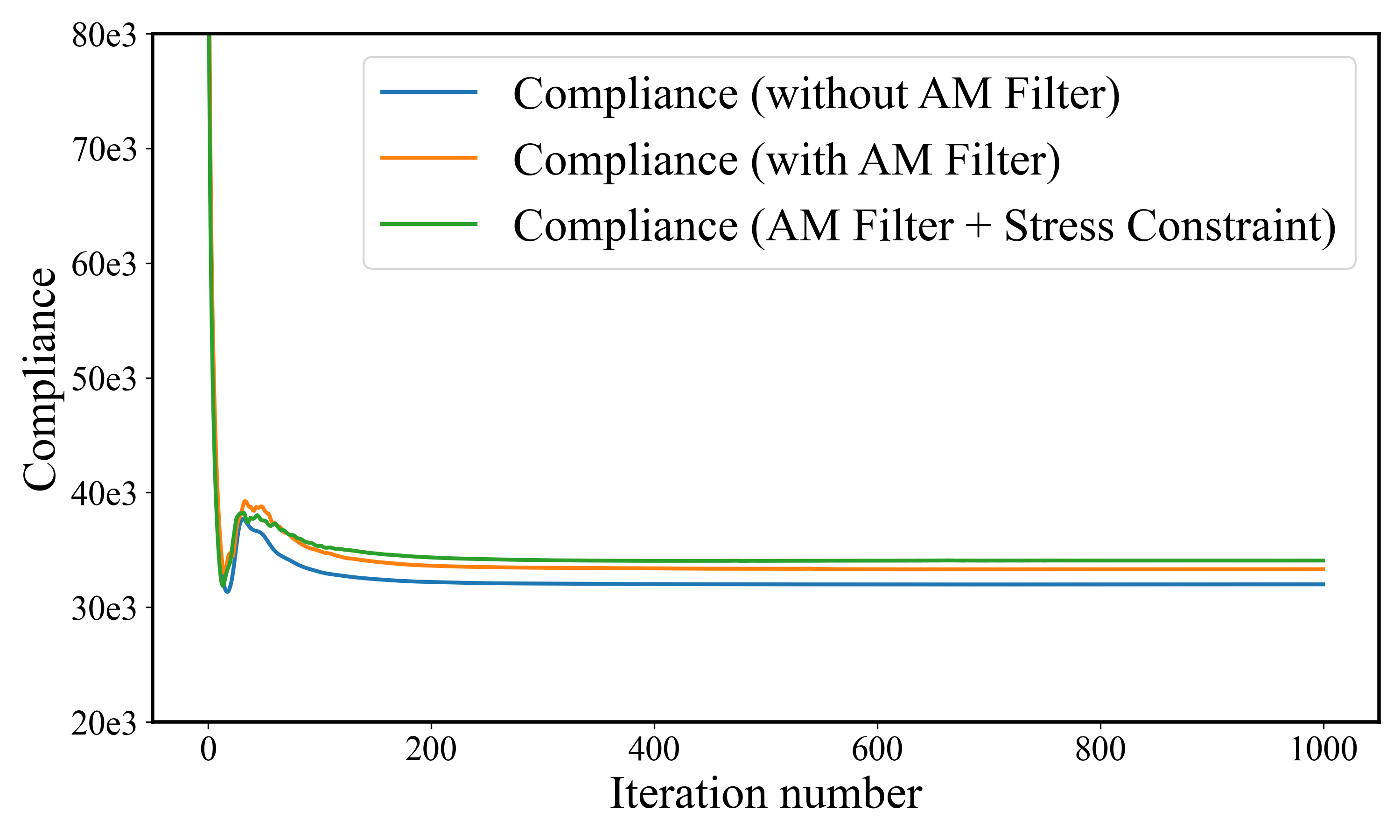} 
	\qquad	
\caption{Comparison of compliance all three conditions .}
    \label{compariosn1}
\end{figure}


This section presents benchmark design problems to evaluate the performance of the proposed framework and highlight the effectiveness of the graph convolutional network in topology optimization. In all cases, material properties are normalized by setting the Young’s modulus to \( E = 1 \)  and the Poisson’s ratio to \( \nu = 0.3 \). To ensure efficient message passing while preserving local structural context, the Chebyshev polynomial degree is set to \( K = 1 \), enabling first-order neighborhood interactions with minimal computational overhead.

The test cases aim to minimize structural compliance while enforcing both AM constraints and stress limitations. The AM filter ensures printability, while a global stress constraint is imposed with an allowable von Mises stress threshold of 2.3. This setting illustrates the framework’s capability to produce high-performance, manufacturable designs that are mechanically robust under prescribed loading.

\subsection{Simply supported beam with distributed loading}
This example investigates a simply supported beam subjected to a uniformly distributed load along its bottom edge. The loading and boundary conditions are illustrated in Figure~\ref{simpBeam}. The design domain is discretized using a structured mesh of \(60 \times 20\) bilinear quadrilateral elements. To simulate fixed support regions and prevent material redistribution in specific areas, the bottom-most layer of elements is designated as a passive zone, where the pseudo-density is fixed at \( \rho = 1 \).
The objective of this optimization problem is to minimize structural compliance, subject to a volume fraction constraint of 0.5.

The results are compared across three configurations: (i) without any AM constraints, (ii) with the AM filter only, and (iii) with both the AM filter and stress constraints activated.
\fref{with_stress_constraints}(a) shows the optimized topology obtained without enforcing any manufacturability constraints. The resulting layout consists of horizontal struts and flat overhanging surfaces that are structurally efficient but non-printable using layer-by-layer additive manufacturing techniques. Incorporating the AM filter into the optimization pipeline, as illustrated in \fref{with_stress_constraints} (b), changes the geometry so that horizontal features are replaced by inclined members, resulting in self-supporting structures. These geometric changes reduce unsupported overhangs and produce a topology suitable for additive production.
\fref{with_stress_constraints} (c) shows the final optimized topology considering both the AM filter and stress constraint. This result highlights the framework’s effectiveness in producing designs that are both manufacturable and mechanically robust.

The convergence behavior of the proposed optimization framework, incorporating the AM filter, is illustrated in Figure~\ref{eval_iter_Pnorm}.
The solid black line represents the evolution of compliance, which initially exhibits sharp fluctuations before undergoing a marked decline after around iteration 200, signifying enhanced structural stiffness as the topology evolves. The dashed black line shows the volume fraction, which begins near 0.72 and steadily decreases, reflecting the optimizer's effort to minimize material usage while adhering to volume constraints.
The dotted curve represents the $p$-norm of the von Mises stress, demonstrating that the proposed method successfully satisfies the stress constraint. This convergence trend underscores the framework’s capability to simultaneously achieve material efficiency and mechanical performance.

\begin{figure}[tb]
  \centering
\includegraphics[width = 0.69\linewidth, trim=0 5pt 0 0,clip]{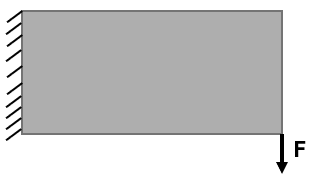}
  \caption{Tip Cantilever.}
    \label{tip_cantilever}
\end{figure}

  


\begin{figure*}[tb]

\centering
  \subfloat[Compliance = 177.94]{\includegraphics[width = 0.3\linewidth, trim=0 5pt 0 0,clip]{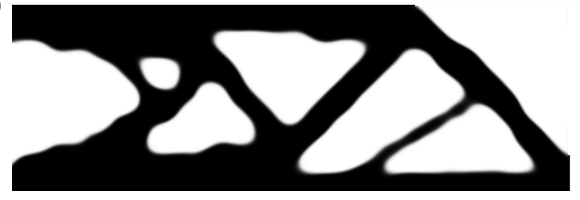}} 
  \hspace{.4cm} 
  \subfloat[Compliance = 187.57]{\includegraphics[width = 0.3\linewidth, trim=0 5pt 0 0,clip]{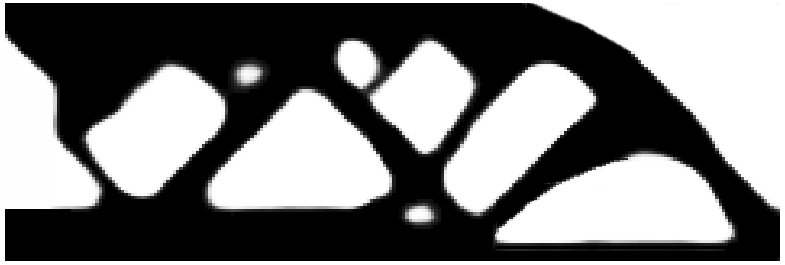}}
  \hspace{.4cm} 
  \subfloat[Compliance = 192.83]
  {\includegraphics[width = 0.3\linewidth, trim=0 5pt 0 0,clip]{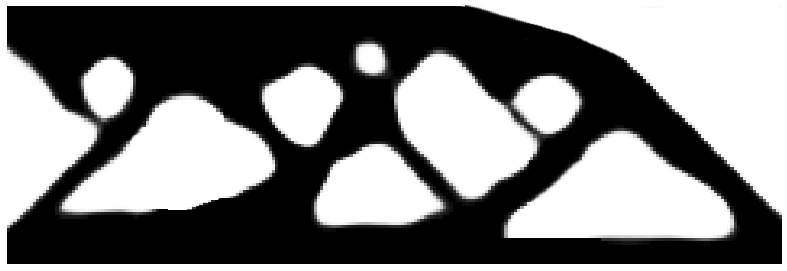}}  
  \caption{Designed topology for tip cantilever. a) without AM filter, b) with AM filter, c) with AM filter + stress constraint.  }
    \label{with_stress_constraints_tip canti}
\end{figure*}
\begin{figure*}[tb!]
	\centering
	\includegraphics[width = 0.9078\linewidth, trim=0 0 0 0,clip]{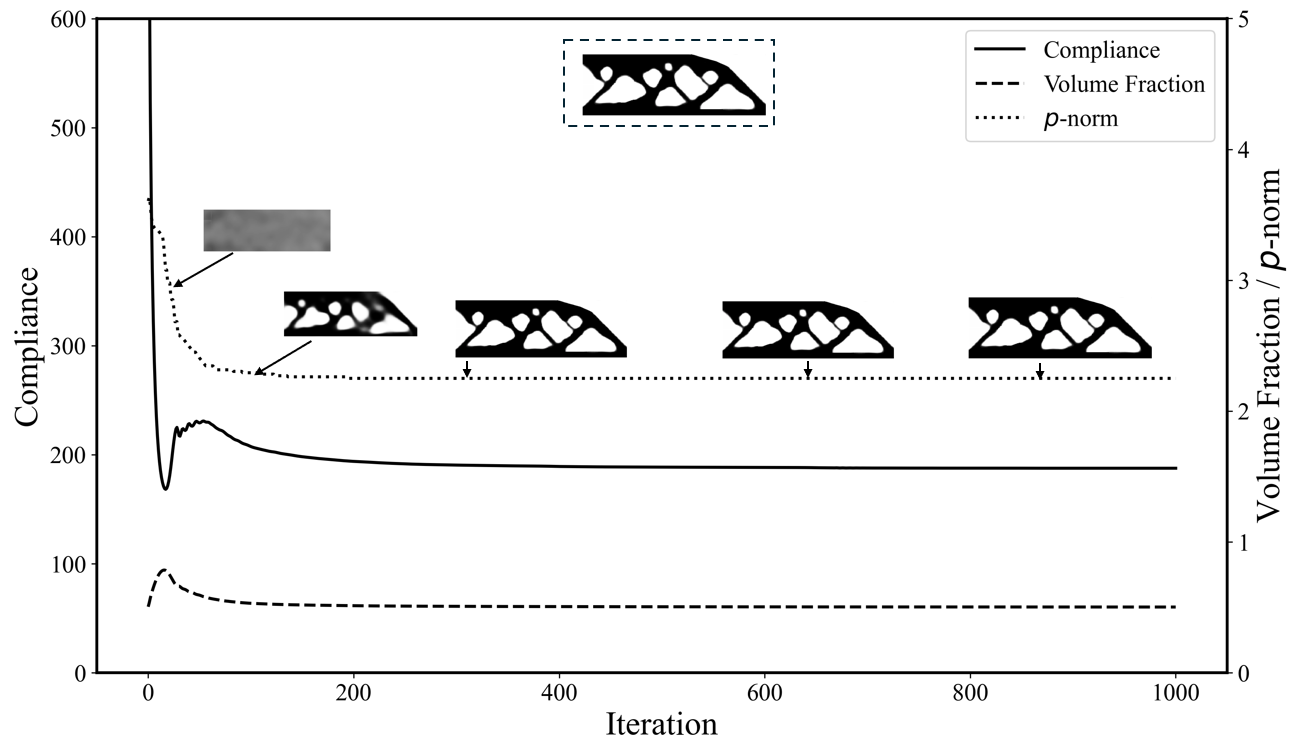} 
	\qquad	
\caption{Evolution of compliance, P-Norm volume fraction over iterations for tip Cantilever Beam .}
    \label{tip_eval_iter_Pnorm}
\end{figure*}

  
\begin{figure}[tb!]
	\centering
	\includegraphics[width = 0.9978\linewidth, trim=0 0 0 0,clip]{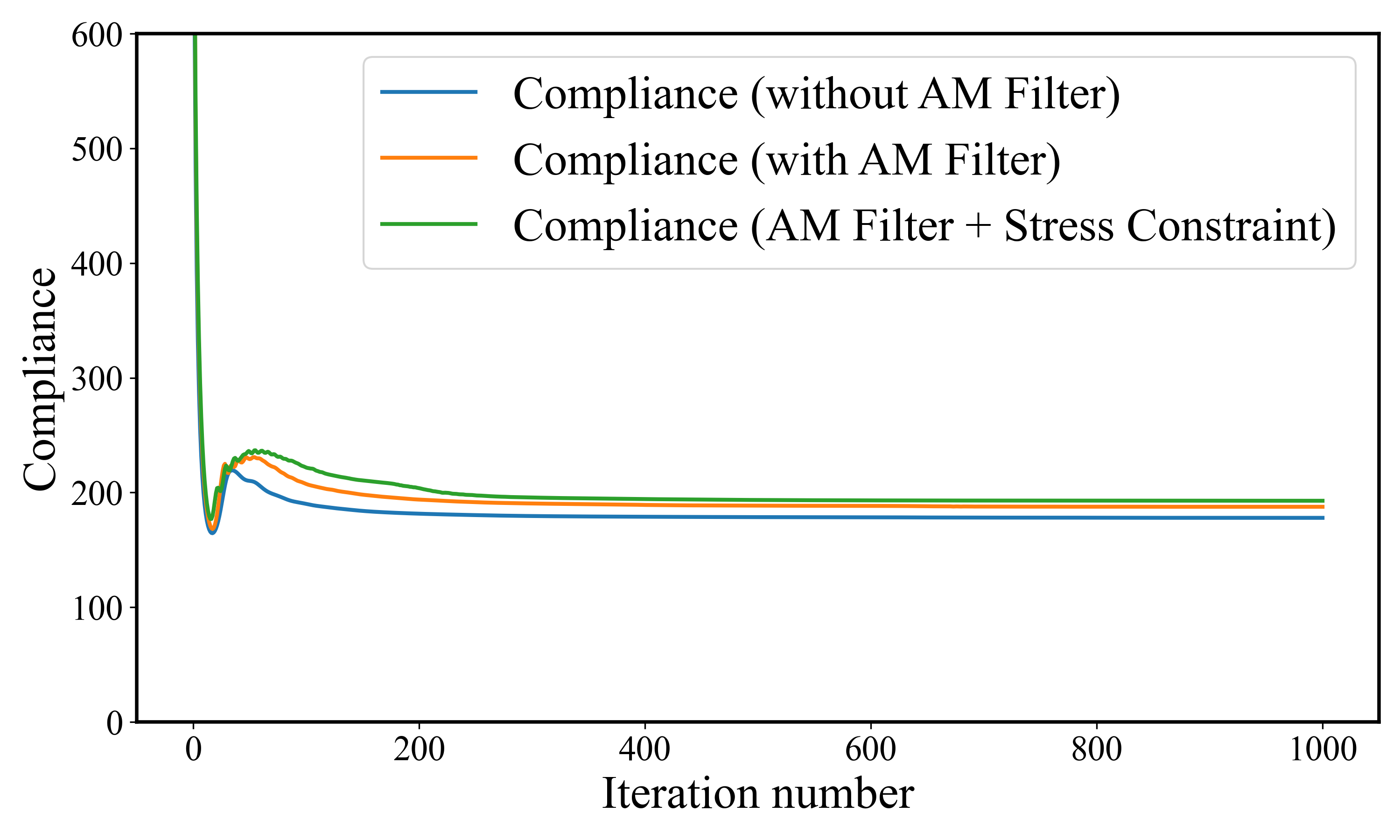} 
	\qquad	
\caption{Comparison of compliance all three conditions for the tip-loaded cantilever beam.}
    \label{compariosntip}
\end{figure}

A comparative analysis of compliance values under the three optimization conditions is provided in \fref{compariosn1}. Under the baseline condition, the optimizer minimizes compliance subject only to a volume constraint, resulting in the lowest compliance values since the solution space remains unconstrained and often yields geometries impractical for additive manufacturing. When the AM overhang filtering is incorporated, the solution is regularized to enforce printability by restricting overhanging features beyond a specified critical angle. Although this introduces geometric constraints that increase compliance, the resulting structures are self-supporting and can be manufactured using standard layer-wise printing methods. The final configuration integrates both the AM filter and global stress constraints via $p$-norm  aggregation of von Mises stress, ensuring the structure remains mechanically reliable, especially in stress-critical regions. Although this scenario produces the highest compliance values among the three, it successfully balances manufacturability and structural performance, yielding designs that are both printable and mechanically robust.

\subsection{Tip-loaded cantilever beam}

This example examines the response of a cantilever beam subjected to a downward load at its free tip, as depicted in \fref{tip_cantilever}. Unlike the simply supported beam, this configuration introduces significant bending moments at the clamped end, providing a challenging scenario to assess the framework’s performance under stress-concentrated regions.

In the absence of any additive manufacturing considerations, the optimization yields a design dominated by extended horizontal features and sharp transitions, as shown in \fref{with_stress_constraints_tip canti}(a). While structurally efficient, these features are incompatible with standard AM practices due to unsupported overhangs and the need for extensive support structures during fabrication.

To address these limitations, the AM overhang filter is introduced, leading to the topology presented in \fref{with_stress_constraints_tip canti}(b). The geometry adapts by favoring diagonal load paths and reducing horizontal overhangs, resulting in a more practical, self-supporting structure. This modification demonstrates the optimizer’s ability to respond to manufacturing constraints while maintaining satisfactory load-bearing capacity.

Building upon this, the third configuration enforces a global stress constraint alongside the AM filter. The resulting design, displayed in \fref{with_stress_constraints_tip canti}(c), achieves a balance between manufacturability and mechanical integrity. Notably, material is redistributed to strengthen regions near the fixed end, where stress concentrations are highest, leading to smoother transitions and improved stress management.

The evolution of compliance and stress performance throughout the optimization process is depicted in \fref{tip_eval_iter_Pnorm}. Initially, the compliance curve exhibits a high value, reflecting the structural inefficiencies present in the early-stage topology. Rapid fluctuations can be observed in the first 100 iterations as the optimizer explores various design configurations and local minima. As the optimization progresses, the compliance stabilizes and gradually decreases, indicating that the optimizer effectively enhances the structure’s load-carrying capacity while satisfying the volume constraint.

The dashed curve represents the volume fraction, which remains relatively constant after the initial iterations, signifying that the optimizer quickly identifies a feasible material distribution that aligns with the specified volume constraint. Meanwhile, the dotted curve captures the evolution of the $p$-norm  of the von Mises stress. Initially, this metric is high, indicating significant stress violations in the design. However, as iterations continue, the \textit{p}-norm value steadily declines and stabilizes around the allowable threshold, demonstrating the optimizer’s success in redistributing material to alleviate stress concentrations and achieve a stress-compliant solution.

The figure includes snapshots of the evolving topology at different iteration stages. These images illustrate the progressive refinement of the structure: from an underdeveloped geometry in the early iterations to a well-defined and efficient design that satisfies both manufacturability and mechanical reliability. The consistent reduction in compliance, coupled with the stabilization of volume fraction and $p$-norm  stress, highlights the framework’s capacity to produce manufacturable, mechanically robust designs through an integrated topology optimization approach.

 The compliance values achieved under the three configurations are compared in  \fref{compariosntip}. As expected, the unconstrained baseline delivers the lowest compliance but fails to meet AM constraints. Introducing the AM filter increases compliance slightly due to geometric limitations, while the stress-constrained solution yields the highest compliance but offers improved reliability under mechanical loads.

\begin{figure}[tb]
  \centering
\includegraphics[width = 0.69\linewidth, trim=0 5pt 0 0,clip]{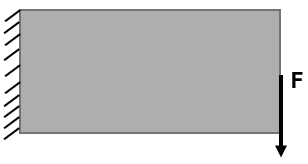}
  \caption{Mid Cantilever.}
    \label{mid_cantilever}
\end{figure}

  

\begin{figure*}[tb]
  \centering
  \subfloat[Compliance = 175.48]{\includegraphics[width = 0.3\linewidth, trim=0 5pt 0 0,clip]{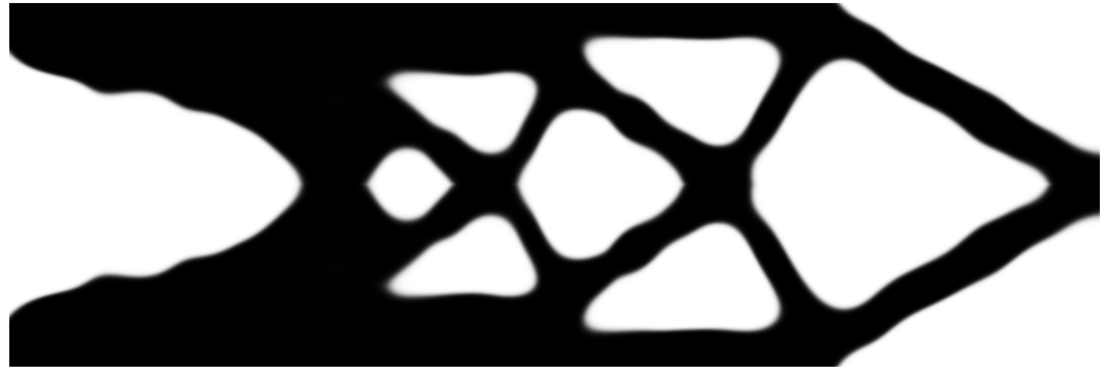}} 
  \hspace{.4cm} 
  \subfloat[Compliance = 188.87]{\includegraphics[width = 0.3\linewidth, trim=0 5pt 0 0,clip]{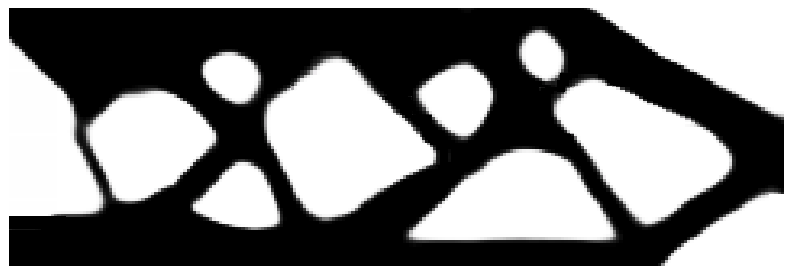}}
  \hspace{.4cm} 
  \subfloat[Compliance = 192.83]
  {\includegraphics[width = 0.3\linewidth, trim=0 5pt 0 0,clip]{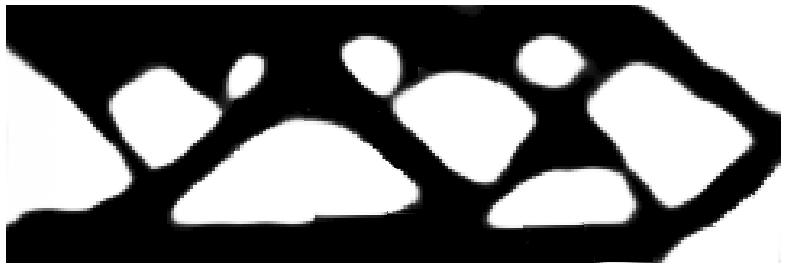}}
  
  \caption{Designed topology for mid cantilever. a) without AM filter, b) with AM filter, c) with AM filter + stress constraint. }
    \label{midcant_withfilter}
\end{figure*}

\begin{figure*}[tb!]
	\centering
	\includegraphics[width = 0.9078\linewidth, trim=0 0 0 0,clip]{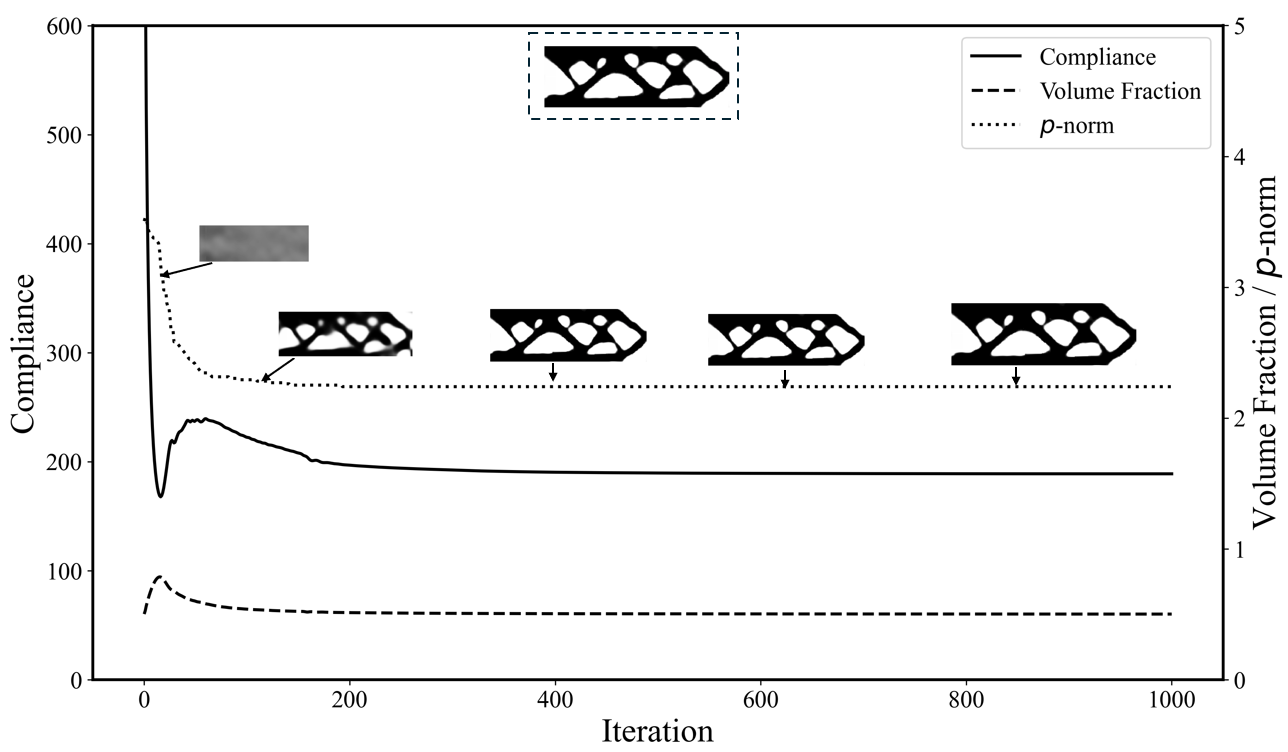} 
	\qquad	
\caption{Evolution of compliance, P-Norm volume fraction over iterations for mid Cantilever Beam .}
    \label{mid_eval_iter_Pnorm}
\end{figure*}

  
\begin{figure}[tb!]
	\centering
	\includegraphics[width = 0.9978\linewidth, trim=0 0 0 0,clip]{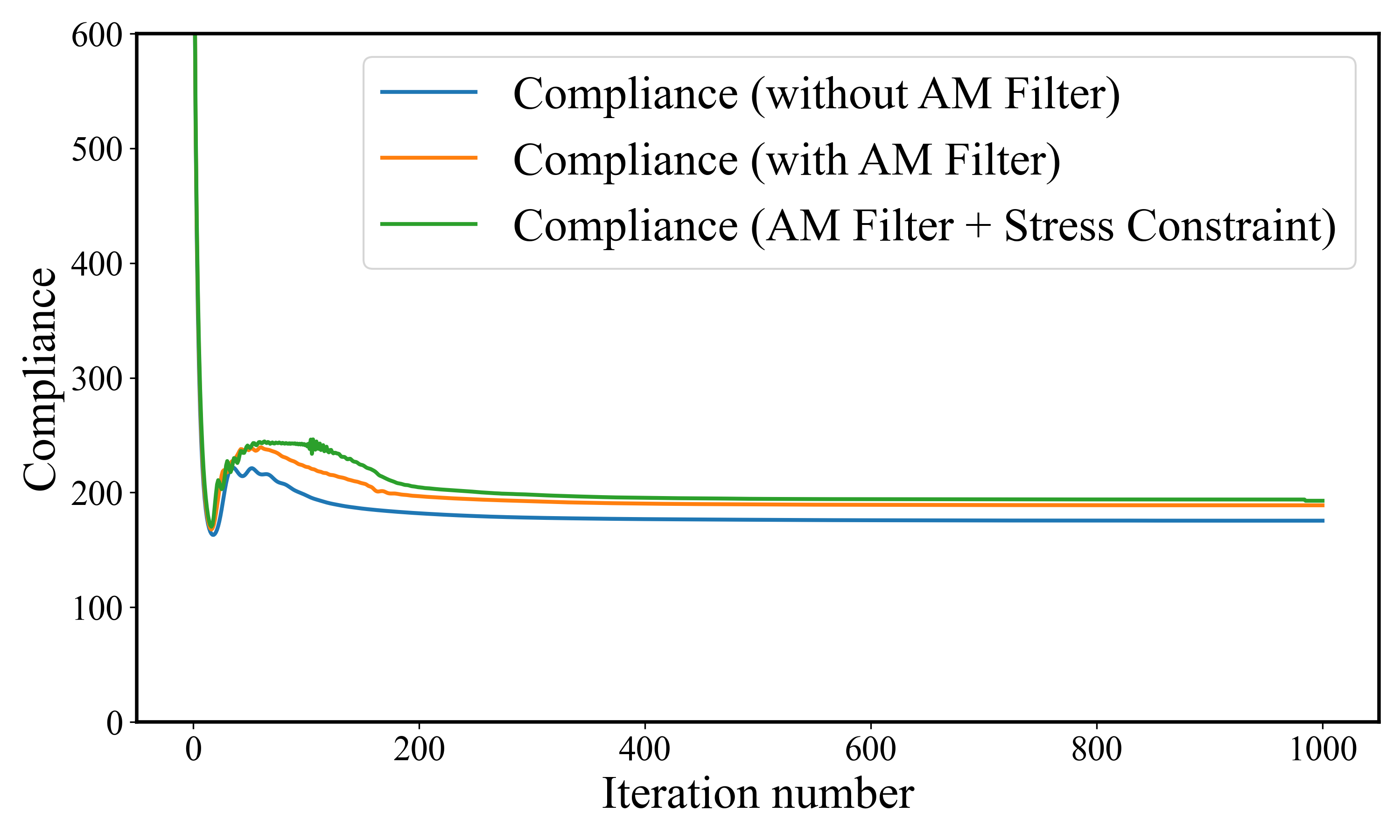} 
	\qquad	
\caption{Comparison of compliance all three conditions for the mid-loaded cantilever beam .}
    \label{compariosnmid}
\end{figure}
\subsection{Midpoint-loaded cantilever beam}
A cantilever beam with a concentrated load at the midpoint of its free edge is considered to test the framework’s ability to generate topologies that respect inherent problem symmetries while accommodating practical design constraints. This configuration naturally suggests a symmetric solution about the horizontal axis due to the balanced load application and geometry.

In the initial unconstrained scenario, the optimizer yields a design that effectively exploits the symmetry of the problem, as seen in \fref{midcant_withfilter}(a). The layout features prominent horizontal members and extended overhangs, characteristic of compliance-driven solutions unconcerned with fabrication limitations. These features, while mechanically efficient, highlight the limitations of unconstrained designs in the context of additive manufacturing.

To promote printability, the AM filter is activated, reshaping the design into a self-supporting configuration with inclined members, as shown in \fref{midcant_withfilter}(b). Interestingly, this filter introduces asymmetries that disrupt the horizontal mirror symmetry of the unconstrained solution, illustrating how manufacturability constraints can influence the natural symmetries of a structure.

Incorporating the stress constraint alongside the AM filter further modifies the design. The resulting topology, depicted in \fref{midcant_withfilter}(c), redistributes material to reinforce stress-prone regions while maintaining overall printability.

The evolution of compliance, volume fraction, and stress performance is presented in \fref{compariosnmid}. The compliance curve shows a rapid decrease during the initial iterations as the optimizer identifies efficient load paths, stabilizing beyond iteration 200. The volume fraction trend indicates progressive adherence to the material constraint, while the $p$-norm  stress metric demonstrates a steady reduction, highlighting the framework’s ability to control stress concentrations. Topology snapshots embedded along the iteration axis illustrate the design’s refinement, evolving from disconnected features to continuous, manufacturable structures. 

The compliance evolution for the mid-loaded cantilever beam under three optimization scenarios is presented in \fref{compariosnmid}. The initial compliance drops sharply as the optimizer identifies efficient load paths, with all cases stabilizing around iteration 200. The unconstrained design achieves the lowest compliance, followed by the AM-filtered design, while the inclusion of the stress constraint leads to the highest compliance values. This trend reflects the trade-offs between mechanical efficiency and the enforcement of manufacturability and stress limitations.


\section{Conclusion}

This study presents a GNN-based framework that effectively serves as a neural field for topology optimization tailored to AM. By treating the design domain as a continuous, learnable representation over a finite element mesh, the method functions as a neural field that predicts material distributions, capturing spatial correlations and design features essential for efficient topology generation. The framework integrates an AM filter to enforce printability by reducing overhangs and ensuring self-supporting structures, while a global stress constraint is implemented using a differentiable $p$-norm of the von Mises stress to maintain mechanical robustness.
A key strength of the approach lies in its fully differentiable design; all modules, including the AM filter and stress evaluation, are embedded within a computational graph that leverages automatic differentiation. This eliminates the need for explicit sensitivity derivation, streamlining the optimization process and enhancing computational efficiency.

Design examples, including simply supported and cantilever beams, demonstrate the framework’s versatility in generating lightweight, manufacturable, and stress-compliant topologies. The neural field nature of the GNN ensures smooth transitions and consistent spatial features, while the differentiable architecture enables seamless integration of constraints and filters. Compared to traditional methods, the proposed approach offers superior automation, adaptability, and practical relevance for AM-ready, structurally sound designs. This research highlights the potential of neural field-based, differentiable frameworks in advancing topology optimization for next-generation manufacturing technologies.
\section*{Statements and Declarations}

\textbf{Funding} \\
This work was supported by the \textit{Institute of Digital Engineering}.

\vspace{0.5em}
\textbf{Conflict of interest}\\ On behalf of all authors, the corresponding author states that there is no conflict of interest.

\textbf{Replication of results}\\ The machine learning models, training scripts, and topology optimization code used in this study are available from the corresponding author upon reasonable request.

\vspace{0.5em}
\textbf{Ethical Approval} \\
The authors state that this article does not contain any studies with human participants or animals performed by any of the authors.


\vspace{1em}
\textbf{Consent for Publication} \\
The authors declare that they all agree to publish this manuscript.

\vspace{0.5em}
\textbf{Author Contributions} \\
Alireza Tabarraei: Conceptualization, Methodology, Supervision, Funding acquisition, Software, Formal analysis, Visualization, Writing of the original draft, review and editing. 
Saquib Ahmad Bhuiyan: Methodology, Software, Formal analysis, Visualization, Writing of the original draft.
\bibliographystyle{elsarticle-num}
\bibliography{Bib_Topology_Optimization_NN_Paper}


\appendix

\end{document}